\documentclass{IEEEtran}
\pdfoutput=1
\usepackage[latin1]{inputenc}

\renewcommand{\vec}[1]{\ensuremath{\boldsymbol{#1}}} 
\usepackage[german,english]{babel}
\usepackage[latin1]{inputenc}
\usepackage[pdftex]{graphicx}
\usepackage{eurosym}
\usepackage{pstool}
\usepackage{acronym}
\usepackage{amssymb}

\usepackage{algorithm}
\usepackage{algpseudocode}
\algdef{SE}[DOWHILE]{Do}{doWhile}{\algorithmicdo}[1]{\algorithmicwhile\ #1}%

\usepackage[cmex10]{amsmath}
\usepackage{tabularx}
\newcommand{\PreserveBackslash}[1]{\let\temp=\\#1\let\\=\temp}

\begingroup
\makeatletter
\catcode`\#=11
    \gdef\pstool@bitmap@opts{%
      -dAutoFilterColorImages#false
      -dAutoFilterGrayImages#false %
      -dColorImageFilter#/FlateEncode %
      -dGrayImageFilter#/FlateEncode 
      }
    \gdef\pstool@pspdf@opts{%
      -dPDFSETTINGS#/prepress %
      -dCompatibilityLevel#1.3 %
      -dEmbedAllFonts#true %
      -dSubsetFonts#true
      }
\endgroup
\graphicspath{{figures/}}

\ifCLASSINFOpdf
\else
\fi
\usepackage[cmex10]{amsmath}

\usepackage{eurosym}

\usepackage{siunitx}
\DeclareSIUnit \voltampere { VA } 
\DeclareSIUnit \watthour { Wh } 
\begin{document}
%
\title{Optimal Sizing and Placement of Distributed Storage in Low Voltage Networks}

\author{
\IEEEauthorblockN{Philipp Fortenbacher\\ Martin Zellner \\ Göran Andersson}
\IEEEauthorblockA{ Power Systems Laboratory \\
ETH Zurich\\
Zurich, Switzerland\\
\{fortenbacher, andersson\}@eeh.ee.ethz.ch, zellnerm@student.ethz.ch}
}


\maketitle

\begin{abstract}
This paper proposes a novel algorithm to optimally size and place storage in low voltage (LV) networks based on a linearized multiperiod optimal power flow method which we call forward backward sweep optimal power flow (FBS-OPF). We show that this method has good convergence properties, its solution deviates slightly from the optimum and makes the storage sizing and placement problem tractable for longer investment horizons. We demonstrate the usefulness of our method by assessing the economic viability of distributed and centralized storage in LV grids with a high photovoltaic penetration (PV). As a main result, we quantify that for the CIGRE LV test grid distributed storage configurations are preferable, since they allow for less PV curtailment due to grid constraints.      
\end{abstract}


\acrodef{LV}[LV]{Low Voltage}
\acrodef{AC-OPF}[AC-OPF]{AC Optimal Power Flow}
\acrodef{OPF}[OPF]{Optimal Power Flow}
\acrodef{FBS-OPF}[FBS-OPF]{Forward Backward Sweep Optimal Power Flow}
\acrodef{FBS}[FBS]{Forward Backward Sweep}
\acrodef{IP}[IP]{Interior Point}
\acrodef{LP}[LP]{Linear Programming}
\acrodef{SOCP}[SOCP]{Second Order Cone Programming}
\acrodef{SDP}[SDP]{Semi Definite Programming}

\section{Introduction}
The need of energy storage in power systems has emerged due to the fact that excess energy from intermittent renewable sources (RES) has increased. In the next years, it can be expected that many battery systems will be installed in the \ac{LV} distribution grid to balance the high in-feed of photovoltaics (PV) with the electricity demand. The main advantage of distributed configurations (storage locations at each single household) as compared with centralized storage configurations (storage location at substation) is that it is possible to reduce overvoltages, line overloadings and network losses in the \ac{LV} network by absorbing locally PV power at times of high PV infeed, such that grid expansion or PV curtailment can be mitigated. It can be foreseen that those configurations can achieve higher PV hosting capacities than centralized ones for the same aggregated storage size. In contrast, centralized storage configurations reduce LV network losses when they are used to absorb energy from higher voltage levels and have lower investment costs. 

In this paper we assess which storage configuration is economically preferable if a group of residential customers wants to invest in storage assets and to maximize its self-consumption from their PV assets. For this purpose, we need to solve an optimal storage sizing and placement problem that considers grid and storage constraints. Some recent papers propose methods that do either not consider grid constraints \cite{Bahramirad2012,Harsha2014} or can only be applied on the transmission level \cite{Ghofrani2013,Pandzic2014}. By introducing the grid constraints the problem gets hard to solve, since the incorporated non-linear AC power flow equations make the sizing and placement problem intractable for long investment horizons. To make the problem tractable one can include convex approximations of the AC power flow equations. This is done in \cite{Low2014,Christakou2015}, where the \ac{AC-OPF} can be approximated to a \ac{SOCP} problem, which is solvable in polynomial time, but still hard to solve. The authors of \cite{molzahn} relax the \ac{AC-OPF} to a convex \ac{SDP} problem and ensure optimality under certain assumptions. However, this is even harder to solve than an \ac{SOCP} problem. A linear approximation of the AC power flow equations was reported in \cite{Coffrin2012}, but this approximation does not exploit the structure of a radial system that is mainly present in LV networks.          

The contribution of this paper is two-fold. First, we introduce a method to solve a distributed storage sizing and placement problem that incorporates a linearized version of the AC power flow equations. In fact, we recast the non-linear \ac{AC-OPF} into a \ac{LP} problem by exploiting the radial structure of an LV network. To enhance optimality, our new \ac{OPF} method iteratively solves the \ac{LP} problem by updating the voltages with a combined forward backward sweep load flow \cite{Teng}. Therefore, we call our new \ac{OPF} method \ac{FBS-OPF}. With the \ac{FBS-OPF} we solve a tractable multiperiod optimal power flow problem including additional sizing and placement constraints. 

Secondly, with our novel developed method we assess which storage configuration (centralized or distributed) is more viable under the assumption that a group of costumers has energy market access and PV units in an LV network. As a result, we will determine storage price levels that indicate at which point storage integration is profitable depending on the centralized and distributed scenario.  

The paper includes following parts. Section~\ref{sec:fbsopf} presents the \ac{FBS-OPF} formulation including the linear power flow approximations and a comparison between the \ac{AC-OPF}. Section~\ref{sec:storcon} copes with the optimal storage sizing and placement problem. Section~\ref{sec:casestudy} shows a case study that compares centralized and distributed storage configurations in terms of viability and storage cost. Section~\ref{sec:con} concludes and provides an outlook for future research.

\section{\acf{FBS-OPF}}
\label{sec:fbsopf}
This Section deals with the recast of the non-linear \ac{AC-OPF} problem into an \ac{LP} problem. Based on the \ac{FBS} power flow method from \cite{Teng} we linearly approximate voltage, power losses and branch flow limits as a function of the nodal reactive and active power for a radial network. These approximations are then incorporated into our linear \ac{FBS-OPF} problem. 

In general, we can calculate the per unit complex nodal current injection vector $\underline{\vec{i}} \in \mathbb{R}^{n \times 1}$ by
\begin{equation}
	\underline{\vec{i}} =  \underbrace{\mathrm{diag}\{1/\underline{v}_1,\ldots,1/\underline{v}_n\}^*}_{\underline{\vec{V}}_\mathrm{df}}[\vec{p} + j\vec{q}]^* \quad,  \label{eq:I}
\end{equation}
\noindent where $\vec{p} \in \mathbb{R}^{n \times 1}$ and $\vec{q} \in \mathbb{R}^{n \times 1}$ are the balanced three-phase real and reactive per unit power injections at each bus $n$. The variables $\underline{v}_1,\ldots,\underline{v}_n$ are the complex nodal line to neutral voltages in per unit.  
According to \cite{Teng} we can define a matrix $\vec{M}_\mathrm{f}  \in \mathbb{R}^{l \times n}$ that maps the nodal current injection vector $\underline{\vec{i}}$ to the branch current vector $
\underline{\vec{i}}_\mathrm{b} \in \mathbb{R}^{l \times 1}$ with
\begin{equation}
	\underline{\vec{i}}_\mathrm{b} = \vec{M}_\mathrm{f} \ \underline{\vec{i}} \quad, \label{eq:Ib}
\end{equation}
where $l$ denotes the number of branches in a radial network. The matrix $\vec{M}_\mathrm{f} \in \mathbb{R}^{l \times n}$ is also called bus-injection to branch-current (BIBC) matrix. Here, we also define a reduced version of $\vec{M}_\mathrm{f}$ indicated with $\vec{M} \in \mathbb{R}^{l \times n-1}$, in which the column of the involved slack bus is deleted.
\subsection{Voltage Approximation}
By applying Ohm's law the voltage drops across the lines can be exactly expressed by 
\begin{eqnarray}
\Delta \underline{\vec{v}}  & = & \vec{M}^T[\vec{R}_\mathrm{d}+j\vec{X}_\mathrm{d}]\vec{M}_\mathrm{f}{\underline{\vec{V}}}_\mathrm{df}^*[\vec{p} + j\vec{q}]^* \quad, \label{eq:compV}
\end{eqnarray}
\noindent where $\vec{R}_\mathrm{d} = \mathrm{diag}\{r_{\mathrm{d}1},...,r_{\mathrm{d}l} \} \in \mathbb{R}^{l \times l}$ is the branch resistance matrix in per unit and $\vec{X}_\mathrm{d}= \mathrm{diag}\{x_{\mathrm{d}1},...,x_{\mathrm{d}l} \} \in \mathbb{R}^{l \times l}$ is the reactance matrix in per unit.
However, \eqref{eq:compV} is complex, such that we have to find a linear and real approximation in reactive and active power for our linear optimization problem. If we assume that the nodal voltage angles are small ($<10^{\circ}$) and the $R/X$ ratio is high ($>2$), which is usually the case for \ac{LV} networks, we can approximate \eqref{eq:compV} to absolute voltage drops with respect to the slack bus voltage $\vec{v}_\mathrm{s} \in \mathbb{R}^{l \times 1}$ by
\begin{equation}
 \vec{v} - \vec{v}_\mathrm{s} \approx \underbrace{\left[ \vec{M}^T\vec{R}_\mathrm{d}\vec{M}_\mathrm{f}|\underline{\vec{V}}_\mathrm{df}| \quad \vec{M}^T\vec{X}_\mathrm{d}\vec{M}_\mathrm{f}|\underline{\vec{V}}_\mathrm{df}| \right]}_{\vec{B}_\mathrm{v}}\left[\begin{array}{c} \vec{p}  \\
 \vec{q}  \end{array} \right] \label{eq:approxV} .
\end{equation}
\subsection{Loss Approximation}
\begin{figure}[!t]
	\centering
	\psfragfig[width = \columnwidth]{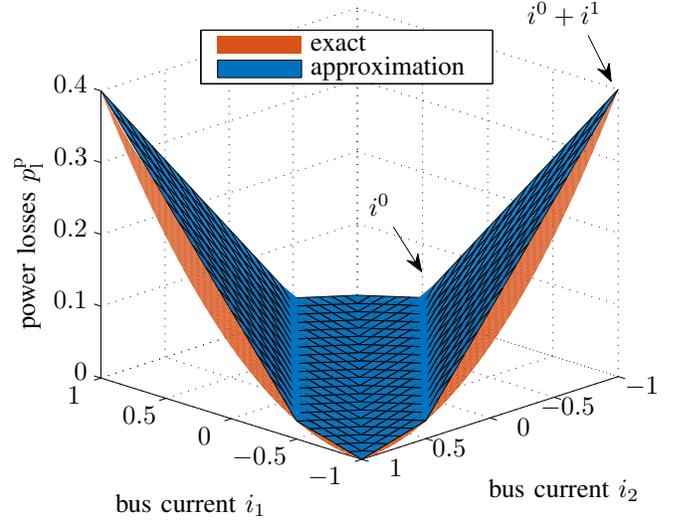}
	\caption{Example of the piecewise linear loss approximation for a two-bus system with one line. The blue surface approximates the exact losses (red surface) by defining four planes.}
	\label{fig:losses}
\end{figure}

We can approximate the losses by
\begin{eqnarray}
\vec{p}_\mathrm{l} &= &\vec{R}_\mathrm{d} \underline{\vec{i}}_\mathrm{b}\circ\underline{\vec{i}}_\mathrm{b}^* \quad,\\
				  & \approx &\underbrace{\vec{R}_\mathrm{d} (\vec{M}_\mathrm{f}|\underline{\vec{V}}_\mathrm{df}|\vec{p})^2}_{\vec{p}_\mathrm{l}^\mathrm{p}} + \underbrace{\vec{R}_\mathrm{d}(\vec{M}_\mathrm{f}|\underline{\vec{V}}_\mathrm{df}|\vec{q})^2}_{\vec{p}_\mathrm{l}^\mathrm{q}}. \label{eq:exactlosses}
\end{eqnarray}
Note that we also assume here that the voltage angles are small to neglect the contribution of the voltage's imaginary part. To find linear expressions in $\vec{p}$ and $\vec{q}$, we approximate their quadratic functions into piecewise linear functions. For this purpose, we divide the quadratic functions into four regions that are defined by the branch currents $\vec{i}^0$ and $\vec{i}^1$ and the branch resistances. We thus obtain following convex formulation for the power losses that are induced by the active bus power injections 
\begin{equation}
\vec{p}_\mathrm{l}^\mathrm{p} \approx \max\left\{\vec{L}_0\vec{p},-\vec{L}_0\vec{p},\vec{L}_1\vec{p} +\vec{b},-\vec{L}_1\vec{p} +\vec{b} \right\}\quad, \label{eq:pwalosses}
\end{equation}
\noindent where 
\begin{eqnarray}
\vec{L}_0 &= &\mathrm{diag}\{i^0_{1},\hdots,i^0_l \} \vec{R}_\mathrm{d}\vec{M}_\mathrm{f}|\underline{\vec{V}}_\mathrm{df}| \quad, \label{eq:plane1} \\
 \vec{L}_1 & =&\mathrm{diag}\{i^0_{1}+i^1_{1},\hdots,i^0_l+i^1_l\}\vec{R}_\mathrm{d}\vec{M}_\mathrm{f}|\underline{\vec{V}}_\mathrm{df}| \quad,\\
 \vec{b} &= &-[r_{\mathrm{d}1} i^0_{1}i^1_{1},\hdots, r_{\mathrm{d}l} i^0_{l}i^1_{l}]^T \label{eq:plane3} \quad. 
 \end{eqnarray}
Equations \eqref{eq:plane1}-\eqref{eq:plane3} define hyperplanes for the power losses. As an example, we show the loss approximation for a two bus system with one line in Fig.~\ref{fig:losses}. The red surface is the exact solution of \eqref{eq:exactlosses}. By defining four planes with supporting points $\pm i^0$ and $\pm(i^0+i^1)$ the blue surface represents the piecewise linear approximation \eqref{eq:pwalosses}.
 In the same straightforward way, we can express the losses caused by the reactive bus power by
\begin{equation}
\vec{p}_\mathrm{l}^\mathrm{q} \approx \max\left\{\vec{L}_0\vec{q},-\vec{L}_0\vec{q},\vec{L}_1\vec{q} +\vec{b},-\vec{L}_1\vec{q} +\vec{b} \right\} \label{eq:epiq} \quad.
\end{equation}
\subsection{Branch Flow Approximation}
The branch flow currents are exactly given by
\begin{equation}
\underline{\vec{i}}_\mathrm{b} = \vec{M}_\mathrm{f}\underline{\vec{V}}_\mathrm{df}^*[\vec{p} + j\vec{q}]^* \label{eq:branchex}\quad.
\end{equation}
If we assume that the reactive power injections are much smaller than the active power injections, which holds for normal grid operation in \ac{LV} grids, we can neglect the contribution on the reactive power by approximating \eqref{eq:branchex} into
\begin{equation}
\vec{i}_\mathrm{b} \approx \underbrace{\vec{M}_\mathrm{f}|\underline{\vec{V}}_\mathrm{df}|}_{\vec{B}_\mathrm{r}}\vec{p}  \quad. \label{eq:branch}
\end{equation}

\subsection{\ac{FBS-OPF} Formulation}
With the approximations from the previous Sections the \ac{AC-OPF} problem can be recasted into a computationally less complex problem. We define following optimization vector $\vec{x} = \left[\vec{p}_\mathrm{gen},\vec{q}_\mathrm{gen},\vec{p}_\mathrm{l}^\mathrm{p},\vec{p}_\mathrm{l}^\mathrm{q},\vec{v} \right]^T$. If we consider positive linear prices $\vec{c}_\mathrm{p}$ for the active generator powers the \ac{OPF} problem can be approximated to the following \ac{LP} problem:
\begin{equation}
\begin{array}{lllll}
  \multicolumn{5}{l}{J(|\underline{\vec{v}}|)^*=\displaystyle\min_{\vec{x}} \vec{c}_\mathrm{p}^T \vec{p}_\mathrm{gen} } \\
\hspace{0.0cm}\text{s.t.}    
& \text{(a)} & \multicolumn{3}{l}{\vec{1}^T\vec{C}_\mathrm{g}\vec{p}_\mathrm{gen} - \vec{1}^T\vec{p}_\mathrm{l}^\mathrm{p}-\vec{1}^T\vec{p}_\mathrm{l}^\mathrm{q}  = \vec{1}^T\vec{p}_\mathrm{d}  }\\
& \text{(b)} & \multicolumn{3}{l}{\vec{B}_\mathrm{v} \left[\begin{array}{l} \vec{C}_\mathrm{g}\vec{p}_\mathrm{gen} \\ \vec{C}_\mathrm{g}\vec{q}_\mathrm{gen}  \end{array} \right] -\vec{v}  = \vec{B}_\mathrm{v}\left[\begin{array}{l} \vec{p}_\mathrm{d} \\ \vec{q}_\mathrm{d}  \end{array} \right] - \vec{v}_\mathrm{s} } \\
& \text{(c)} & \multicolumn{3}{l}{\vec{p}_\mathrm{l}^\mathrm{p} - \vec{L}_0\vec{C}_\mathrm{g}\vec{p}_\mathrm{gen} \geq -\vec{L}_0\vec{p}_\mathrm{d} } \\
& \text{(d)} & \multicolumn{3}{l}{\vec{p}_\mathrm{l}^\mathrm{p} + \vec{L}_0\vec{C}_\mathrm{g}\vec{p}_\mathrm{gen} \geq \vec{L}_0\vec{p}_\mathrm{d} } \\
& \text{(e)} & \multicolumn{3}{l}{\vec{p}_\mathrm{l}^\mathrm{p} - \vec{L}_1\vec{C}_\mathrm{g}\vec{p}_\mathrm{gen} \geq -\vec{L}_1\vec{p}_\mathrm{d} + \vec{b} } \\
& \text{(f)} & \multicolumn{3}{l}{\vec{p}_\mathrm{l}^\mathrm{p} + \vec{L}_1\vec{C}_\mathrm{g}\vec{p}_\mathrm{gen} \geq +\vec{L}_1\vec{p}_\mathrm{d} + \vec{b} } \\
& \text{(g)} & \multicolumn{3}{l}{\vec{p}_\mathrm{l}^\mathrm{q} - \vec{L}_0\vec{C}_\mathrm{g}\vec{q}_\mathrm{gen} \geq -\vec{L}_0\vec{q}_\mathrm{d} } \\
& \text{(h)} & \multicolumn{3}{l}{\vec{p}_\mathrm{l}^\mathrm{q} + \vec{L}_0\vec{C}_\mathrm{g}\vec{q}_\mathrm{gen} \geq \vec{L}_0\vec{q}_\mathrm{d} } \\
& \text{(i)} & \multicolumn{3}{l}{\vec{p}_\mathrm{l}^\mathrm{q} - \vec{L}_1\vec{C}_\mathrm{g}\vec{q}_\mathrm{gen} \geq -\vec{L}_1\vec{q}_\mathrm{d} + \vec{b} } \\
& \text{(j)} & \multicolumn{3}{l}{\vec{p}_\mathrm{l}^\mathrm{q} + \vec{L}_1\vec{C}_\mathrm{g}\vec{q}_\mathrm{gen} \geq +\vec{L}_1\vec{q}_\mathrm{d} + \vec{b} } \\
& \text{(k)} & \multicolumn{3}{l}{-\vec{i}_\mathrm{b}^\mathrm{max} + \vec{B}_\mathrm{r}\vec{p}_\mathrm{d}\leq \vec{B}_\mathrm{r}\vec{C}_\mathrm{g}\vec{p}_\mathrm{gen} \leq \vec{i}_\mathrm{b}^\mathrm{max} + \vec{B}_\mathrm{r}\vec{p}_\mathrm{d}}\\
& \text{(l)} &\vec{v}_\mathrm{min} \leq \vec{v} \leq \vec{v}_\mathrm{max} \\
& \text{(m)} &\vec{p}_\mathrm{min} \leq \vec{p}_\mathrm{gen} \leq \vec{p}_\mathrm{max} \\
& \text{(n)} &\vec{q}_\mathrm{min} \leq \vec{q}_\mathrm{gen} \leq \vec{q}_\mathrm{max} \quad,
\end{array}
\label{eq:FBOPF}
\end{equation}
\noindent where $\vec{p}_\mathrm{d} \in \mathbb{R}^{n \times 1}$ and $\vec{q}_\mathrm{d} \in \mathbb{R}^{n \times 1}$ are the active and reactive load consumption for $n$ buses. Further, the active and reactive $n_\mathrm{g}$ generator bus injections $\vec{p}_\mathrm{gen} \in \mathbb{R}^{n_\mathrm{g} \times 1}$ and $\vec{q}_\mathrm{gen} \in \mathbb{R}^{n_\mathrm{g}\times 1 }$  are mapped to the buses with the Matrix $\vec{C}_\mathrm{g} \in \mathbb{R}^{n \times n_\mathrm{g}}$. Equation (\ref{eq:FBOPF}a) specifies the power balance in the grid. The voltage approximation \eqref{eq:approxV} is included in (\ref{eq:FBOPF}b). The constraints (\ref{eq:FBOPF}c-j) incorporate  epigraph formulations of \eqref{eq:pwalosses} and \eqref{eq:epiq} that approximate the power losses. Since \eqref{eq:pwalosses} and \eqref{eq:epiq} are convex, we convert the $\max$ operator into equivalent inequalities. Note to get an optimal solution the solver can only select values for $\vec{p}_\mathrm{l}^\mathrm{p}$ and $\vec{p}_\mathrm{l}^\mathrm{q}$ that lie on the defined hyperplanes. Constraint (\ref{eq:FBOPF}k) include branch flow limits, and the constraints (\ref{eq:FBOPF}l-n) specify the lower and upper bounds for the voltage, active and reactive generator powers. In this formulation, we consider boxed-bounded active and reactive power settings. However, we could also include a piecewise-linear approximation to allow for a specific power factor range as reported in \cite{forte_powertech}.   
\subsection{\acf{FBS} Algorithm}
The matrices $\vec{B}_\mathrm{v}$,$\vec{B}_\mathrm{r},\vec{L}_0$, and $\vec{L}_1$ depend on the nodal voltage magnitudes $|{\vec{v}}|$. By setting the initial voltages to $\vec{1}$ we overestimate the voltage drops and branch flows to some extent, which leads to the fact that the optimal solution deviates slightly. However, we can reduce this error if we iteratively solve the optimization problem with the \ac{FBS} power flow algorithm, described as Algorithm~\ref{algo} below.
In step~\ref{start} we initialize the start voltages with the slack bus voltages. In addition, we calculate the hyperplane parameters for two operating currents. The maximum current on the branches we expect is $\vec{i}^0+\vec{i}^1=\vec{M}_\mathrm{f}\vec{C}_\mathrm{g}\vec{p}_\mathrm{max}$. The second region we define is at 25\% of this maximum current.  After solving the problem (step~\ref{problem}) in stage $h$, we calculate in the forward stage the currents (step~\ref{forward}). Then, the voltages are updated in the backward stage at step~\ref{backward}. These steps are iteratively repeated until the mean absolute error is below a predefined threshold $\epsilon$  (step~\ref{break}). Since forward/backward sweep methods have a high convergence rate, typically $h$ is small ($h\leq4$) \cite{Teng}.
\begin{algorithm}[h!]
	\begin{algorithmic}[1]
		\State {$\underline{\vec{v}}^0 = \vec{v}_\mathrm{s}, h = 0, \vec{i}^0 = 0.25 \vec{M}_\mathrm{f}\vec{C}_\mathrm{g}\vec{p}_\mathrm{max}, $ }
	    \Statex $\vec{i}^1 = 0.75 \vec{M}_\mathrm{f}\vec{C}_\mathrm{g}\vec{p}_\mathrm{max}$ \label{start}
		\Do  
		\State $ \vec{x} = \min \ J(|\underline{\vec{v}}^h|)$ \label{problem}
		\State $\underline{\vec{i}}^h = {\mathrm{diag}\{\underline{\vec{v}}^h \}}^{*-1} [(\vec{C}_\mathrm{g}\vec{p}_\mathrm{gen}-\vec{p}_\mathrm{d}) - j (\vec{C}_\mathrm{g}\vec{q}_\mathrm{gen}-\vec{q}_\mathrm{d})]$ \label{forward}
		\State $\underline{\vec{v}}^{h+1} = \vec{v}_\mathrm{s} + \vec{M}^T[\vec{R}_\mathrm{d}+j\vec{X}_\mathrm{d}]\vec{M}_\mathrm{f}\underline{\vec{i}}^h$  \label{backward}
		\State $h = h+1$
		\doWhile{$\mathrm{mean}(|\vec{v}^{h-1} - \vec{v}^h|) > \epsilon$} \label{break}
		\vspace{-0.1cm}
	\end{algorithmic}
	\caption{FBS-OPF algorithm.}
	\label{algo}
\end{algorithm}

\subsection{Multiperiod \ac{FBS-OPF}}
With the introduction of storage the \ac{FBS-OPF} problem \eqref{eq:FBOPF} is coupled in time, since the energy can be transferred from one time step to another. As a consequence we need to extend the single step OPF problem into a multiperiod \ac{OPF} problem. To do so, let $\vec{X} = [\vec{x}_1,\cdots,\vec{x}_N]^T$ the new optimization vector, then we can specify following multiperiod problem over a given horizon $N$:
\begin{equation}
	\begin{array}{lllll}
		\displaystyle \min_{\vec{X}} 
		&  \multicolumn{4}{l}{T \sum\limits_{k=0}^{N}\vec{c}_\mathrm{p}^T(k) \vec{p}_\mathrm{gen}(k) } \\
		\hspace{0.0cm}\text{s.t.} & \multicolumn{4}{l}{\forall \vec{x}: (\ref{eq:FBOPF}\mathrm{a})-(\ref{eq:FBOPF}\mathrm{n})} \quad,
	\end{array}
	\label{eq:mFBOPF} 
\end{equation}
\noindent where $T$ is the sample time of a time-variable price profile $\vec{c}_\mathrm{p}(k)$. Depending on the energy product $T$ can range from minutes (intraday trading) to hours (spotmarket, base products).

\subsection{Comparison between \ac{AC-OPF} and \ac{FBS-OPF}}
We aim to compare our \ac{FBS-OPF} method with a standard \ac{AC-OPF} in terms of convergence and optimality. In addition, we also compare the execution times between the aforementioned methods in our case study (s. Section~\ref{sec:casestudy}). 
\subsubsection{Testsystem}
\begin{figure}[!t]
	\centering
    \includegraphics[width = \columnwidth]{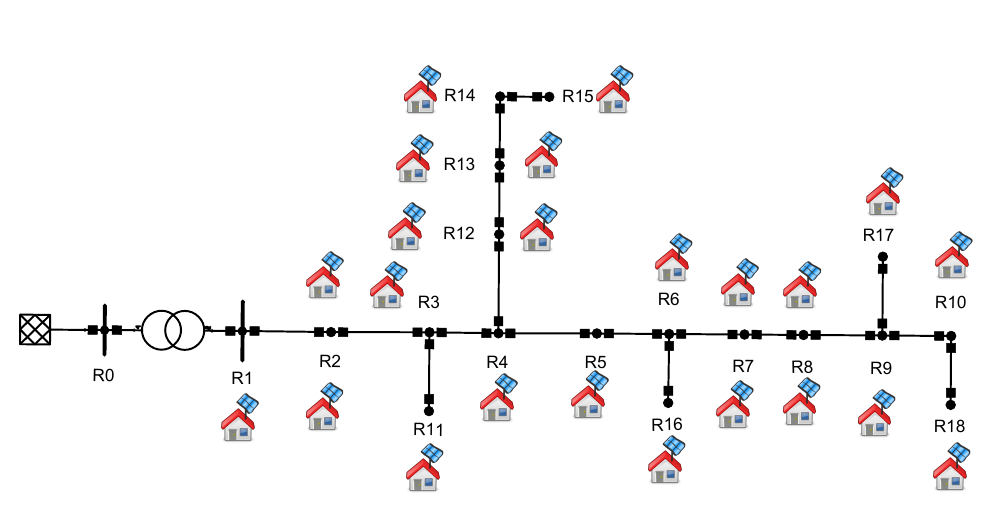}
    \caption{CIGRE test grid from \cite{cigre} configured with a high PV infeed penetration.}
  \label{fig:cigregrid}
\end{figure}

For our comparison, we use the European CIGRE \ac{LV} benchmark grid \cite{cigre} which is depicted on Fig.~\ref{fig:cigregrid}. We solve problem \eqref{eq:FBOPF} with the parameters that are listed in Table~\ref{tab:scenario}.  For this configuration, the solver tries to maximize the PV infeed while satisfying voltage and line constraints.
\begin{table}[!t]
	\caption{Parameters for OPF comparison.}
	\centering
	\begin{tabular}{lll}
		\hline
		Load & $p_{\mathrm{d}1},...,p_{\mathrm{d}18}$ & 5kW \\
		     &$q_{\mathrm{d}1},...,q_{\mathrm{d}18}$ & 1kVar \\
		\hline
		PV Generators &$p_{\mathrm{gen}1}^{\mathrm{max}},...,p_{\mathrm{gen}18}^{\mathrm{max}}$ & 30kW \\
					  &$p_{\mathrm{gen}1}^{\mathrm{min}},...,p_{\mathrm{gen}18}^{\mathrm{min}}$ & 0kW \\
				      &$q_{\mathrm{gen}1}^{\mathrm{max}},...,q_{\mathrm{gen}18}^{\mathrm{max}}$ & 10kVar \\
					  &$q_{\mathrm{gen}1}^{\mathrm{min}},...,q_{\mathrm{gen}18}^{\mathrm{min}}$ & -10kVar \\
					  &$c_{\mathrm{p}1},...,c_{\mathrm{p}18}$ & 20\euro/kWh \\
					 \hline
		Feeder 		 &$p_{\mathrm{gen}0}^{\mathrm{max}}$ & 1MW \\
					 	&$p_{\mathrm{gen}0}^{\mathrm{min}}$ & 1MW \\
					 	&$q_{\mathrm{gen}0}^{\mathrm{max}}$ & 1MVar \\
					 	&$_{\mathrm{gen}0}^{\mathrm{min}}$  & -1MVar \\
					 	&$c_{\mathrm{p}0}$ & 30\euro/kWh \\
		\hline
		Line Parameters & from \cite{cigre} \\
		\hline
	\end{tabular}
	\label{tab:scenario}
	\vspace{-0.5cm}
\end{table}
\subsubsection{Convergence}
\begin{figure}[!t]
	\centering
    \psfragfig[width = \columnwidth]{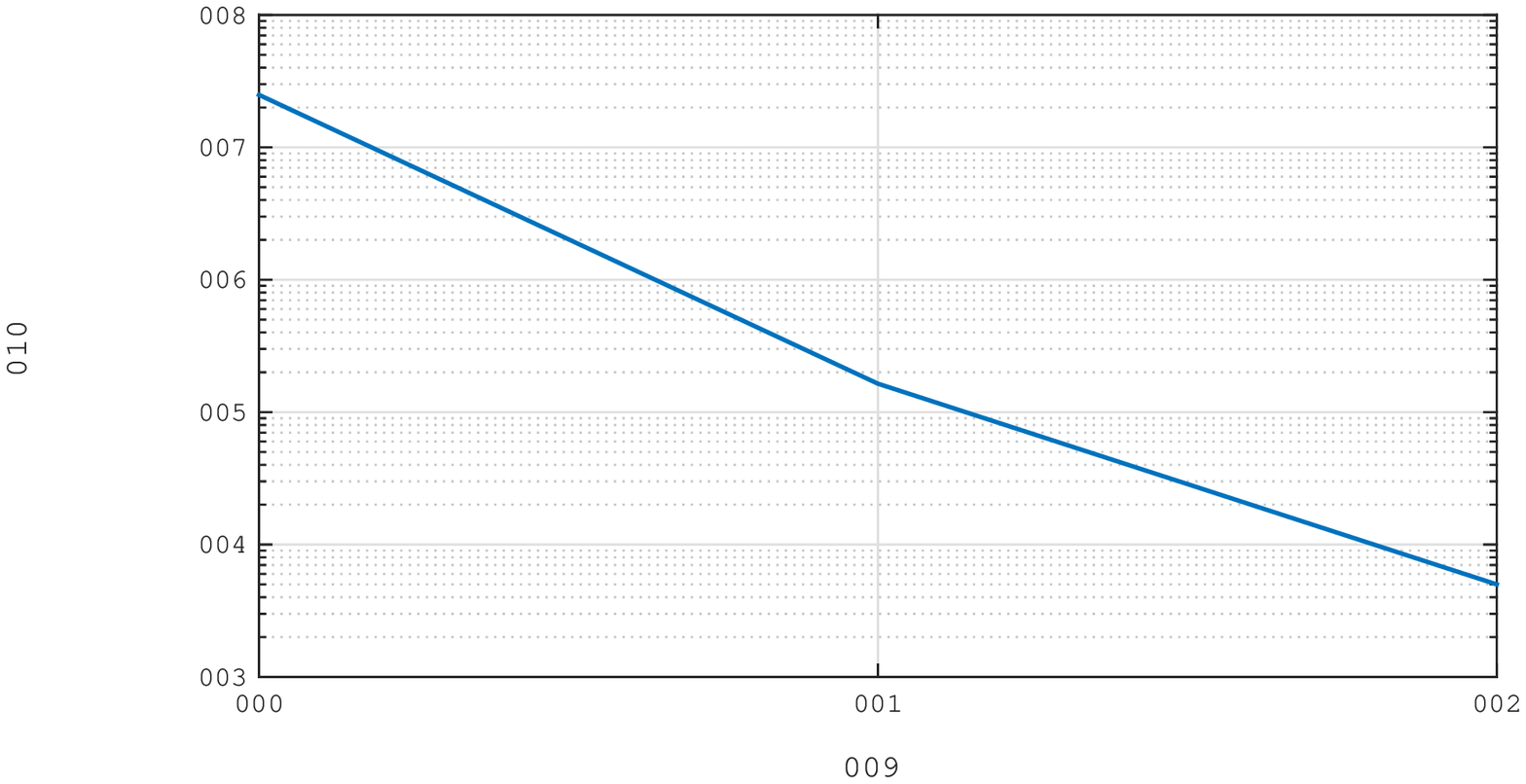}
    \caption{Voltage convergence rate of the \acf{FBS-OPF} as a function of the iteration $h$.}
  \label{fig:convergence}
\end{figure}

At first we investigate the convergence rate of our proposed \ac{FBS-OPF} algorithm. Figure~\ref{fig:convergence} shows the voltage mean absolute error (MAE) as a function of the iteration number $h$. To compute the MAE, we take the difference between the voltages ($\vec{v}$) from the optimization \eqref{eq:FBOPF} and the correct voltage values that were found by running a power flow with the generator setpoints $\vec{p}_\mathrm{gen},\vec{q}_\mathrm{gen}$ from the optimization \eqref{eq:FBOPF}. As a result, the voltages from the optimization are higher than the correct ones. This means that we obtain voltage projections that always lie within the voltage limits. For one iteration ($h=1$) the voltage error already constitutes 2.5e-3 p.u., which is small related to possible network parameter inaccuracies. Due to these findings we think that one iteration is reasonable for sizing and placement of storage.      
\subsubsection{Optimality}
\begin{figure}[!t]
	\centering
    \psfragfig[width = \columnwidth]{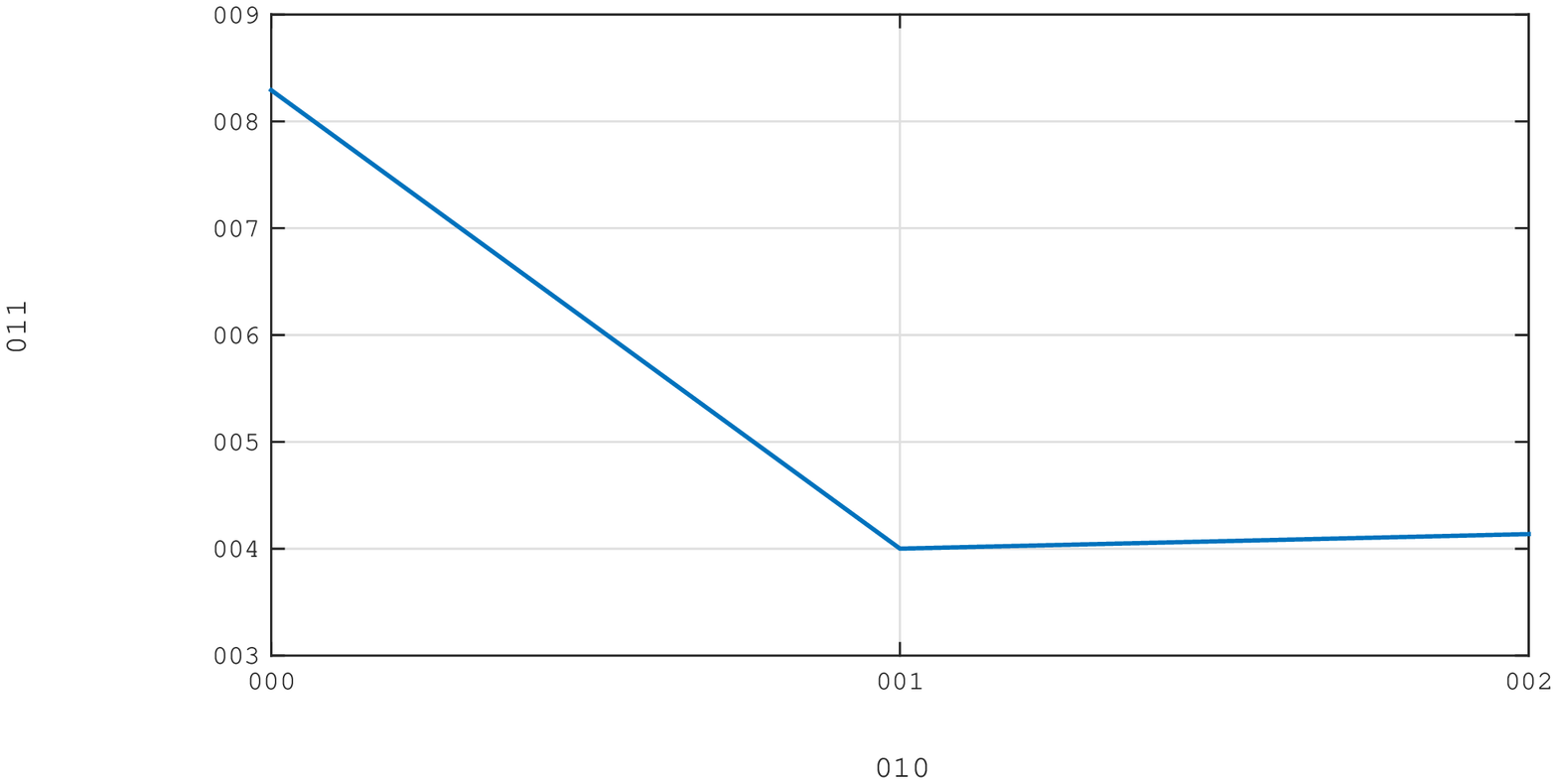}
    \caption{\acf{FBS-OPF} objective value error related to the \acf{AC-OPF} objective value.}
  \label{fig:optimality}
\end{figure}

Due to its non-linear constraints the \ac{AC-OPF} problem is known to be non-convex. The non-linear \ac{AC-OPF} problem can be solved with an \ac{IP} method \cite{matpower}. However, this solving technique does not guarantee an optimal solution, which means that it is difficult to assess optimality with regard to an \ac{IP} method. But it is still worthwhile to compare the objective values of the \ac{AC-OPF} with the \ac{FBS-OPF}, since this gives us a better insight on the performance of the \ac{FBS-OPF}. Figure~\ref{fig:optimality} shows the objective value error of the \ac{FBS-OPF} related to the \ac{AC-OPF}. We can observe that at the first iteration the \ac{FBS-OPF} value is 2\% above the \ac{AC-OPF} value, while for higher iteration numbers the \ac{FBS-OPF} value is smaller. The smaller value can be explained by that the branch flow limit at the feeder is slightly violated due to the approximation \eqref{eq:branch} meaning more apparent power is exported by the feeder than allowed. Still we observe, by using the one iteration approximation, the problem is feasible in the constraints, but slightly diverges to the exact solution. Again, we think that this is reasonable for sizing and placement strategies, since such strategies imply long planning horizons that are also subject to uncertainties.

\section{Intertemporal coupled multi-period \ac{OPF}}
\label{sec:storcon}
In this Section we focus on how we incorporate battery storage to the presented multiperiod \ac{FBS-OPF} problem \eqref{eq:mFBOPF}. This enables us to formulate a storage sizing and placement problem.   
\subsection{Incorporation of Storage}
Storage introduces an intertemporal coupling to the multiperiod problem, since the available storage energy can be delivered within certain time intervals. We can express the time-varying energy level $\vec{e}$ of $n_\mathrm{s}$ storages with following discrete state space equation
\begin{equation}
\vec{e}(k+1)  = \vec{I}\vec{e}(k) + 
\vec{B} \underbrace{\left[\begin{array}{l} \vec{p}_\mathrm{gen}^\mathrm{s,dis} \\ \vec{p}_\mathrm{gen}^\mathrm{s,ch} \end{array}\right]}_{\vec{p}_\mathrm{gen}^\mathrm{s}} \quad,									
\end{equation}
\noindent where $\vec{p}_\mathrm{gen}^\mathrm{s,dis},\vec{p}_\mathrm{gen}^\mathrm{s,ch} \in \vec{p}_\mathrm{gen}$ represent the discharging and charging powers of the storages. The input matrix $\vec{B}$ is
\begin{equation}
\vec{B} =  T [-\mathrm{diag}\{\eta_{\mathrm{dis},1}^{-1},...,\eta_{\mathrm{dis},n_\mathrm{s}}^{-1}\}  \ -\mathrm{diag}\{\eta_{\mathrm{ch},1},...,\eta_{\mathrm{ch},n_\mathrm{s}}\} ] \quad,
\end{equation}
\noindent where $\eta_{\mathrm{ch},i},\eta_{\mathrm{dis},i}$ are the charging and discharging efficiencies. To incorporate the complete energy level  evolution $\vec{E} = [\vec{e}(1),...,\vec{e}(N)]^T$ we can write
\begin{equation}
\vec{E} =  \underbrace{\left[ \begin{array}{c} \vec{I} \\ \vdots \\ \vec{I}\end{array} \right]}_{\vec{S}_x}\vec{e}_0 + \underbrace{\left[\begin{array}{ccc}  \vec{B}  &  & \vec{0} \\ \vdots &\ddots \\ \vec{B} & \cdots & \vec{B} \end{array}\right]}_{\vec{S}_u}\underbrace{\left[\begin{array}{l} \vec{p}_\mathrm{gen}^\mathrm{s}(0) \\ \vdots \\  \vec{p}_\mathrm{gen}^\mathrm{s}(N) \end{array} \right]}_{\vec{U}} \quad, \label{eq:evo}
\end{equation}
\noindent where $\vec{e}_0$ is the initial energy level vector. Consequently, we can define with \eqref{eq:evo} two following inequalities for the multi-period optimization problem
\begin{equation}
\vec{e}_{\mathrm{min}} -\vec{S}_x\vec{e}_0 \leq \vec{S}_u {\vec{U}} \leq \vec{e}_{\mathrm{max}} - \vec{S}_x\vec{e}_0 \label{eq:batcon}\quad,	
\end{equation}
\noindent to operate the storages between the allowed storage level limits $\vec{e}_{\mathrm{min}}$ and $\vec{e}_{\mathrm{max}} $.

\subsection{Optimal Sizing and Placement of Storage}
Given the results from the previous Section we can now define the storage sizing and placement problem. To do so, we make the maximum storage limit $\vec{e}_{\mathrm{max}} $ variable and transform it to the decision variable $\vec{z}$. If we set $\vec{e}_{\mathrm{min}}$  to $\vec{0}$, we can rewrite \eqref{eq:batcon} into following inequalities    
\begin{align}
 \vec{S}_u \vec{U} &- \left[\vec{1}^{N \times 1}\otimes\mathrm{diag}\{\vec{1}^{n_\mathrm{s}\times1} \} \right]\vec{z}& \leq & -\vec{S}_x\vec{e}_0 \label{con:size1} \quad,\\
 -\vec{S}_u \vec{U} & & \leq &\vec{S}_x\vec{e}_0 \label{con:size2} \quad.
\end{align}
We can now include this result into the problem \eqref{eq:mFBOPF}, such that we get
\begin{equation}
\begin{array}{lllll}
J^{*}&  = & \multicolumn{3}{l} {\displaystyle\min_{\vec{X},\vec{z}} \ T \left(\sum\limits_{k=0}^{N} \ \vec{c}_{p}^T(k) \vec{p}_{\mathrm{gen}}(k) \right) + \ \vec{c}_\mathrm{s}^T\vec{z}} \\
& & \text{s.t.} & \multicolumn{2}{l}{\forall \vec{x}: (\ref{eq:FBOPF}\mathrm{a})-(\ref{eq:FBOPF}\mathrm{n})} \\
& & & \multicolumn{2}{l}{\eqref{con:size1},\eqref{con:size2}} \quad,
\end{array}
\label{eq:placesize} 
\end{equation}
\noindent where $\vec{c}_\mathrm{s}^T$ is the storage cost vector.

\section{Case Study}
\label{sec:casestudy}
To illustrate potential applications of our optimal sizing and placement approach we perform a number of simulations to compare the scenarios of centralized and distributed storage in distribution grids.
We assume that a group of (pro-)sumers has access to the energy market  (selling to the grid and buying from the grid) and tries to minimize its energy bill by dispatching their own PV and storage assets. As a main result, we optimally size and place the storage assets for both configurations in dependence of future storage costs and PV share by using yearly spot market prices \cite{eex}. 
\subsection{Definition}
For our simulations we use the CIGRE testgrid \cite{cigre} with 18 households shown in Fig.~\ref{fig:cigregrid} with a PV share of 100\% for a time period of 31 days. We consider a centralized storage at the R0 bus for the centralized case and storages at the buses R1-R18 for the distributed case. To study the economic viability of battery storage in distribution grids, we compare the two scenarios of a centralized storage that is placed at the feeder bus and a decentralized storage configuration with a storage at every single node. The parameters of the two scenarios are shown in Table~\ref{tab:sim_parameters}.
\begin{table}[t]
	\centering
	\caption{Simulation parameters for storage cost analysis.}
	\begin{tabular}{lcc} \hline
		& Centralized & Distributed\\
		\hline
		Storage Node & 0 & 1-18 \\
		Storage Apparent Power & 180 kVA & 10 kVA  \\
		Storage Power Factor &  \multicolumn{2}{c}{0-1 box bounded}  \\
		PV Share & \multicolumn{2}{c}{100 \% ($\hat=$ 18)} \\
		PV Power & \multicolumn{2}{c}{30 kW} \\
		Storage Cost & \multicolumn{2}{c}{ 25- 300 \euro/kWh} \\
		Battery Calendar Life & \multicolumn{2}{c}{10 years} \\
		Storage Charging Efficiency & \multicolumn{2}{c}{88\%} \\
		Storage Discharging Efficiency & \multicolumn{2}{c}{88\%}  \\ 
		Simulation Horizon & \multicolumn{2}{c}{31 \si{\day}}\\
		Time-steps/Resolution & \multicolumn{2}{c} {744@1h}\\
		Busses & \multicolumn{2}{c}{19}\\
		Energy Price Profile & \multicolumn{2}{c}{EEX Spotmarket \cite{eex}} \\
		PV Energy produced &  \multicolumn{2}{c}{147.13 MWh} \\
		Consumed Energy &  \multicolumn{2}{c}{4.84 MWh} \\
		Households & \multicolumn{2}{c}{18} \\
		\hline
	\end{tabular}
	\label{tab:sim_parameters}
\end{table}
\subsection{Implementation}
To perform the simulations a MATLAB simulation framework based on MATPOWER \cite{matpower} was created. The non-linear \ac{AC-OPF} problems were solved using the solver Pardiso \cite{schenk} bundled with IPOPT \cite{waechter} and for the \ac{FBS-OPF} problem we use the CPLEX \cite{cplex} and GUROBI \cite{gurobi} solver. For generating the consumer load profiles, the load profile generator developed in \cite{bucher} was used. For the \ac{FBS-OPF} algorithm a MATPOWER extension was developed. 

\subsection{Computation Time}
With the introduction of storage we gain the possibility to take energy from a certain point in time and release it in another time-step which introduces a coupling between the different time-steps. Unfortunately due to the coupling the problem we have to solve also increases in complexity by $N \cdot n$ buses.

\begin{figure}[t]
\psfragfig[width = \columnwidth]{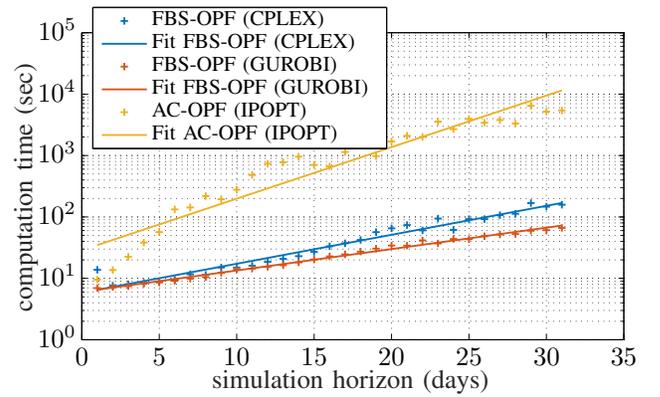}
\caption{Recorded computation time as a function of the simulation horizon and the corresponding lin-log regression fits.}
\label{fig:simulation_time}
\end{figure}

The computation time depends on the simulation horizon $N$, which is shown in Fig.~\ref{fig:simulation_time} for the \ac{FBS-OPF} and the non-linear \ac{AC-OPF}. We observed that the \ac{FBS-OPF} and \ac{AC-OPF} grow approximately polynomially with the computation time. However, the \ac{AC-OPF} problem grows with a higher exponent. 
Due to this property, the introduced \ac{FBS-OPF} algorithm enables us to perform studies with significantly larger time horizons. While for the conventional \ac{AC-OPF} approach, the simulation time is limited by the size of the problem to a couple of days, the \ac{FBS-OPF} allows us to perform simulations for an entire year.

With the \ac{FBS-OPF}, the non-linear \ac{AC-OPF} problem has been simplified to an \ac{LP} problem that can be solved using industry grade \ac{LP}-solvers such as CPLEX \cite{cplex} and GUROBI \cite{gurobi}. Standard solving methods for \ac{LP} problems are the dual simplex method and the \ac{IP} method.
While the \ac{IP} method assures polynomial runtime, the simplex algorithm can have exponential runtime for degenerated problems. For similar problems such as the storage sizing for different prices, the simplex algorithm can however be initialized with the solution of the similar problem and therefore can have better convergence properties. Taking the convergence properties of the \ac{LP} problem of the \ac{FBS-OPF} into account, we can only assure polynomial runtime as an upper boundary when using the \ac{IP} method to solve the \ac{LP} problem. 

\subsection{Economic Viability}
\begin{figure}[t]
	\centering
	\psfragfig[width = \columnwidth]{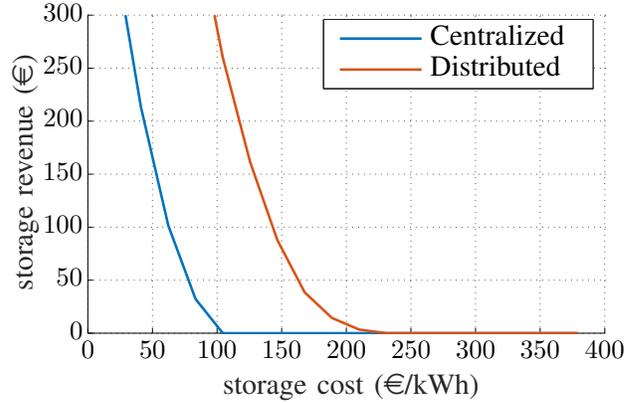}
	\caption{Storage revenue as a function of the storage cost.}
	\label{fig:storage_revenue_price}
\end{figure}

The results from the economic assessment are shown in Fig.~\ref{fig:storage_revenue_price}. Since the simulation horizon is smaller than the expected battery lifetime, we transform the storage cost from the simulation horizon to their expected calendar life. We define the storage revenue as the objective value difference between a configuration with and without storage. For a centralized scenario we can identify a cost of $\approx $100 \euro/kWh, at which a storage installation becomes profitable. While for the distributed scenario, this point is at $\approx $ 230 \euro/kWh. This can be explained by the less utilization of PV energy in the centralized case, since excess PV energy has to be curtailed due to line and transformer overloading and/ or voltage violations. 

\begin{figure}[t]
	\centering
	\psfragfig[width = \columnwidth]{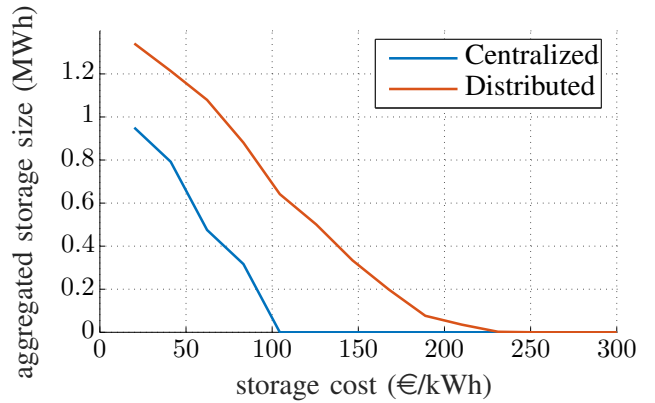}
	\caption{Storage size as a function of the storage cost for the distributed and centralized storage configuration.}
	\label{fig:storage_size_price}
\end{figure}

It becomes also evident that the distributed configuration is generally superior in terms of profit to the centralized solution. On the one hand, this means that distributed configurations achieve profitability for higher storage cost. On the other hand, storage cost for centralized configurations might be lower than for distributed ones. But we think that the cost for centralized storage will not be lower by factor 2 than for distributed storage in future \cite{Schoenung}. Hence, it can be anticipated that distributed configurations will be in this regard economically preferable.
Observing Fig.~\ref{fig:storage_size_price} we again find the profitable points as described before. Analyzing the optimization result for the storage size we can see that the storage size for the distributed storage is higher than for the centralized configuration.

\subsection{Placement of Distributed Storage}
\begin{figure}[t]
	\centering
    \psfragfig[width = \columnwidth]{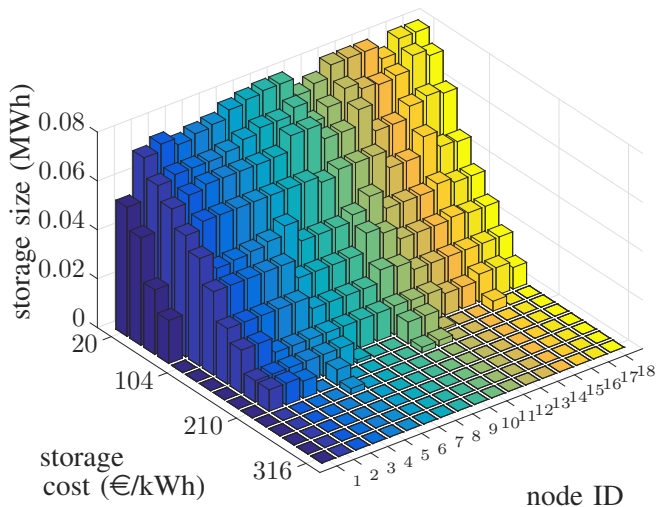}
\caption{Storage placement for the distributed storage configuration as a function of storage investment costs.}
\label{fig:storage_placement}
\end{figure}

Simulating a distributed scenario with more than one storage we get implicitly information about the storage placement within the network. Optimizing the storage size of several storages will as well optimize the distribution of the total storage capacity within the network.
Hence not only the question of the optimal size, but also the question where to place a storage unit of which size is implicitly solved by the optimization. Figure~\ref{fig:storage_placement} shows a placement for the given testgrid shown in Fig.~\ref{fig:cigregrid}. For this grid thermal line constraints are more dominant than voltage constraints. This means that the storage at node 1 is not selective to reduce PV curtailment, since the allowable transformer overloading is higher than for the line between node 1 and 2. For higher storage cost it is better to place the storages rather in the middle of the root of the grid (2,3,4,6) than at the end of the branches. This can be explained that this placement allows for less PV curtailment, but also saves network losses to increase the profit of energy arbitrage on the market.   

\section{Conclusion}
\label{sec:con}
In this paper we present a novel \ac{OPF} method for radial networks that approximates the \ac{AC-OPF} problem to an \ac{LP} problem. This \ac{FBS-OPF} iteratively solves the \ac{LP} problem with a combined forward backward sweep load flow. For the first iteration we show that this new \ac{OPF} has a low error in optimality and is feasible in the voltage constraints. Our approximative method can be incorporated into an optimal sizing and placement problem for distributed storage. We demonstrate the usefulness of our method in a case study where we assess the viability of distributed and centralized storage configurations. For the studied test system it turns out that distributed storage configurations are preferable, since the economic impact of curtailing PV energy is higher than the saved network losses from the energy market transactions. It remains to be investigated how general this conclusion is, but it can be anticipated that for typical LV grid configurations it holds. We show that the \ac{AC-OPF} based optimal sizing and placement problem is intractable for long time horizons, while the \ac{FBS-OPF} based problem can solve yearly investment horizons in reasonable time. However, to study the applicability of our approach for bigger networks with a higher number of storages, it can be foreseen that the problem gets too complex due to the storage coupling and therefore also intractable. Hence, future work relates to further decompose the sizing and placement problem by using distributed optimization techniques. Furthermore, in this paper we consider calendar battery life for the economic assessment, but battery wear also depends on the operational management. Therefore, we also aim to incorporate a battery degradation model in our sizing and placement problem to further investigate the impact of battery degradation on the profitability.




\bibliographystyle{IEEEtran}
\bibliography{literature}

%

\end{document}